\documentclass[sigconf]{acmart}

\usepackage{amsmath}
\usepackage{graphicx}
\usepackage{color}
\usepackage{xspace}

\usepackage{subcaption}

\newcommand{\todo}[1]{{\color{red} TODO: {#1}}}
\usepackage{multirow}
\usepackage{colortbl}
\usepackage[inline]{enumitem}
\usepackage{framed}
\definecolor{shadecolor}{gray}{0.92} 

\definecolor{ao}{rgb}{0.0, 0.5, 0.0}
\def\cca#1{\cellcolor{ao!#1}\ifdim #1pt>49pt\color{white}\fi{#1}}
\definecolor{mygr}{rgb}{0.0, 0.5, 0.0}
\definecolor{mybl}{rgb}{0.0, 0.0, 0.5}
\definecolor{myban}{rgb}{0.5, 0.0, 0.5}
\def\ccb#1{\cellcolor{mybl!#1}\ifdim #1pt>50pt\color{white}\fi{#1}}
\def\ccg#1{\cellcolor{mygr!#1}\ifdim #1pt>50pt\color{white}\fi{#1}}
\def\cca#1{\cellcolor{myban!#1}\ifdim #1pt>45pt\color{white}\fi{#1}}

\newcommand{\ccaTest}[2]{%
  \ifnum #1>40
    \cellcolor{green!#1}%
  \else
    \ifnum #1>20
      \cellcolor{yellow!#1}%
    \else
      \cellcolor{red!#1}%
    \fi
  \fi
  \ifnum #2>49\color{white}\fi #1%
}

\newcommand{\precision}{\% Resolved\xspace}
\newcommand{\swelite}{\texttt{SWE-Bench Lite}\xspace}
\newcommand{\sweverif}{\texttt{SWE-Bench Verified}\xspace}
\newcommand{\shortlite}{\texttt{Lite}\xspace}
\newcommand{\shortverif}{\texttt{Verified}\xspace}
\newcommand{\swebench}{\texttt{SWE-Bench}\xspace}

\newcommand{\approach}[1]{\texttt{#1}}

\newcommand{\llm}[1]{\texttt{#1}}

\newcommand{\totalsubmitters}{77\xspace}

\newcommand{\totalentries}{212\xspace}

\newcommand{\rqevolution}{How have the number of submissions and their precision evolved across the leaderboards over time?}

\newcommand{\rqprofile}{What are the different submitter types represented in the \swebench{} leaderboard?}

\newcommand{\rqproducts}{What types of software products and availability modes are associated with submissions?}

\newcommand{\rqopensource}{How do open-source and closed-source solutions compare?}

\newcommand{\rqLLM}{What large language models, whether used individually or in combination, are employed in submissions?}

\title{What’s in a Benchmark? The Case of SWE-Bench in Automated Program Repair}

\author{Matias Martinez}
\affiliation{%
  \institution{Universitat Politècnica de Catalunya}
  \city{Barcelona}
  \country{Spain}}
\email{matias.martinez@upc.edu}

\author{Xavier Franch}
\affiliation{%
  \institution{Universitat Politècnica de Catalunya}
  \city{Barcelona}
  \country{Spain}}
\email{xavier.franch@upc.edu}

\begin{abstract}
The rapid progress in Automated Program Repair (APR) has been fueled by advances in AI, particularly large language models (LLMs) and agent-based systems. 
SWE-Bench is a  benchmark designed to evaluate  repair systems using real issues mined from popular open-source Python repositories. 
Its public leaderboards—SWE-Bench Lite and Verified—have become central platforms for tracking progress and comparing solutions.
In this paper, we present the first comprehensive study of these two leaderboards, examining who is submitting solutions, the products behind the submissions, the LLMs employed, and the openness of the approaches. 
We analyze 79 entries submitted to Lite leaderboard and 133 to Verified. Our results show that most entries on both leaderboards originate from industry, particularly small companies and large publicly traded companies. 
These submissions often achieve top results, although academic contributions—typically open source—also remain competitive. 
We also find a clear dominance of proprietary LLMs, especially Claude family, with state-of-the-art results on both leaderboards currently achieved by Claude 4 Sonnet. 
These findings offer insights into the SWE-Bench ecosystem that can guide greater transparency and diversity in future benchmark-driven research.
\end{abstract}

\copyrightyear{2026}
\acmYear{2026}
\setcopyright{cc}
\setcctype{by}
\acmConference[ICSE-SEIP '26]{2026 IEEE/ACM 48th International Conference on Software Engineering}{April 12--18, 2026}{Rio de Janeiro, Brazil}
\acmBooktitle{2026 IEEE/ACM 48th International Conference on Software Engineering (ICSE-SEIP '26), April 12--18, 2026, Rio de Janeiro, Brazil}
\acmPrice{}
\acmDOI{10.1145/3786583.3786904}
\acmISBN{979-8-4007-2426-8/2026/04}


\begin{document}

\maketitle

\pagestyle{plain}

\section{Introduction}

The field of Automated Program Repair (APR) has progressed rapidly over the last decade. Early approaches relied on techniques such as search-based methods (e.g., \cite{LeGoues2011genprog}) and constraint-based methods (e.g., \cite{nguyen2013Semfix,Xuan2017nopol}).
More recently, advances in Artificial Intelligence (AI), particularly in deep learning and transformer architectures, have significantly influenced the field. Many current state-of-the-art techniques \cite{ruan2024specrovercodeintentextraction, ma2025alibabalingmaagent, bouzenia2024RepairAgent} are based on agents powered by large language models (LLMs), which can autonomously handle the entire repair process, from understanding the issue to validating the generated patch.
This progress has been significantly facilitated by the creation of bug benchmarks, such as Defects4J \cite{Just2014Defects4J}, which serve as a common ground for evaluating and comparing the performance of APR techniques.

%
%
Recently, new benchmarks have emerged to address evolving scenarios, most notably \swebench{}~\cite{jimenez2024SWEBenchLLMs}, which includes 2,294 issue repair tasks mined from pull requests and issues across 12 widely used Python repositories.
\swebench{} differs from classical APR benchmarks (e.g. \cite{Madeiral2019bears,lin20217Quixbugs}) in several important ways. 
First, each instance in \swebench{} does not come with test cases or scripts that expose the bug and validate candidate fixes. This setup is more aligned with real-world scenarios that developers typically face, where test cases are not available when a bug is reported.
Second, \swebench{} provides a framework for running experiments and computing metrics such as precision (percentage of resolved issues) and project-level repair statistics, similar to what ~\cite{Durieux:2019:RepairThemAll} built on top of Defects4J.
Finally, the creators of \swebench{} actively maintain several public leaderboards where evaluation results are published.
Researchers and developers can submit their results via pull requests. 
Such leaderboards were uncommon in earlier APR research, which often lacked platforms for sharing results.

In less than two years, \swebench{} has experienced a rapid growth demonstrated by: (1) its increasing number of entries, (2) the implication of big players both from industry and academia, (3) the huge number of scientific papers citing the benchmark. This high impact motivates us to investigate the benchmark in more depth: what are the dynamics of the leaderboards, and what type of proposals are being submitted. To this end, we present the first systematic study of \swebench{} leaderboards. We analyze and categorize submitted solutions to identify contributors and their origins (academia, industry, or the open-source community). We further examine which LLMs are used, the degree of openness of both the solutions and the models, and how these factors relate to precision.
We focus on two leaderboards hosted on the \swebench{} website:
\begin{enumerate*}[label=(\alph*)]
\item \swelite{}, which consists of 300 repair tasks selected from the full benchmark; and
\item \sweverif{}, which includes 500 tasks curated and verified by OpenAI~\cite{chowdhury2024swebenchverified}.
\end{enumerate*}

To carry out our study, we examine each entry in the \swebench{} leaderboards.
An entry is a submission reporting the results of evaluating an approach on a subset of \swebench{} instances.
For each entry, we collect the available metadata, including the submission pull request, submitter link, and precision score (\precision).
We then search both academic publications and grey literature (e.g., blog posts, LinkedIn posts) to gather additional information about each submission. 
This includes the nature of the submitter (e.g., company or academic institution), any software artifact related to the submission, the LLMs used, and the openness of the solution. %


In total, we analyze 79 entries from the \shortlite{} leaderboard and 133 from the \shortverif{} leaderboard. 
Our results show that, while there is a diversity of submitter types (e.g., academia, industry, and collaborations), the majority of submissions come from industry, particularly from small companies and large publicly traded corporations such as Amazon, IBM, and Google.
We also identified submissions made by individual developers, highlighting how access to powerful language and code models —whether commercial (e.g. GPT 4, Claude 4) or open-source (e.g., Llama, Qwen)— is enabling the creation of sophisticated repair systems.
We also found that the majority of submissions achieving state-of-the-art results rely on proprietary LLMs —most notably, Claude 4 Sonnet.

Being today the most popular benchmark in APR with high participation of industry, we argue that our findings provide valuable information of the state of the art and the practice. At this respect, we discuss critical aspects of \swebench{}, including patch correctness and benchmark evolution, and derive implications for practitioners, researchers, and benchmark or leaderboard builders.

We structure our analysis around five research questions. 
In all cases, we examine both the number of submissions and the achieved precision (\precision{}).

\begin{enumerate}[leftmargin=*, labelsep=1em]
    \item \textbf{RQ1 (Evolution):} \rqevolution{}
    \item \textbf{RQ2 (Submitter Profile):} \rqprofile{}  
    \item \textbf{RQ3 (Products \& Availability):} \rqproducts{}
    \item \textbf{RQ4 (Open Source):} \rqopensource{}
    \item \textbf{RQ5 (LLMs):} \rqLLM{}
\end{enumerate}

The paper continues as follows.
Section \ref{sec:methodology} presents the methodology of our study.
Section \ref{sec:results} presents the results.
Section \ref{sec:discussion} discusses the results and their implications for researchers, practitioners, and benchmark and leaderboard builders.
Section \ref{sec:ttv} presents the threats to validity.
Section \ref{sec:relatedwork} presents the related work.
Section \ref{sec:conclusion} concludes the paper.


\section{Methodology}

\label{sec:methodology}

\subsection{Terminology}

\label{sec:method:terminology}

We introduce below basic terminology specific to \swebench{}.

{\bf Instance}: A GitHub issue repair task in \swebench{}.

{\bf Leaderboard}: The public website that ranks submissions based on their performance on a defined subset of instances (e.g., \shortverif{}).

{\bf Entry}: A record in a leaderboard, consisting of a name, date, \precision{}, and a URL.

{\bf Submitter}: The individual or team responsible for an entry, submitting it via a pull request to the \swebench{} repository.

{\bf Approach}: The configuration or method evaluated on \swebench{}, typically reflected in the entry's name. For example, \approach{SWE-Agent}.



\subsection{Data Collection}
\label{sec:method:inspection}

We present the method for collecting information from two \swebench{} leaderboards: \swelite{} and \sweverif{}.
We decided not to analyse the other two \swebench{}  leaderboards, namely \texttt{Full} and \texttt{Multimodal}, for the following reasons: 
\begin{enumerate*}[label=(\alph*)] \item \texttt{Full}: 
all solutions submitted to \texttt{Full} are also included in, at least, one of the two considered leaderboards, 
\item \texttt{Multimodal}: we restrict our research to language- (either natural or programming) based agents. \end{enumerate*}

\label{sec:method:inspection:datacollection}
To collect data, we inspect \swelite{} and \sweverif{} as follows.
First, we visit each leaderboard page and retrieve from each entry. 
Then, for each entry, we conduct the following steps:
\begin{enumerate*}[label=(\alph*)] \item We visit the listed site to gather valuable information about the approach  and to find any references related to experiments conducted on SWE-Bench.
\item While browsing a page, we manually navigate the entire site and follow any external links that may be useful for our goal (e.g., link to a blog or a paper).
\item We supplement these data with a dedicated Google search to capture additional pages that are not accessible through the website linked to the leaderboard.
The Google query we create is \texttt{"<Name\_Entry> + SWE-Bench"}; in  \texttt{"<Name\_Entry>"}, we remove the date and the model's details.
For example, given the entry \texttt{"Gru(2024-12-08)"}, the name used in the Google search is \texttt{"GRU"}.
\item We inspect the pages obtained with the web search, and store those that describe the approach or the experiment conducted on \swebench{}.
\item We also inspect the data submitted for each entry, notably the \texttt{README.md} 
and \texttt{metadata.yaml} files, available in the \swebench's repository\footnote{\url{https://github.com/SWE-bench/experiments/tree/main/evaluation/}} dedicated to collect the row information from the submissions.
\item We also visit the LinkedIn of the submitter, and eventually some of its members, in order to detect posts that discuss the experiment.

In addition to collecting the submitter information for each entry, we also identify the underlying approach associated with it.
For example, the entries \texttt{"SWE-agent + GPT-4o (2024-05-13)"} and \texttt{"SWE-agent + GPT-4 (1106)"} are based on the same approach, \texttt{SWE-agent}, but differ in the LLM used (GPT-4o and GPT-4, resp.).
\end{enumerate*}

Availability and configuration of data has sometimes restricted the type of analysis. For instance, we initially planned an analysis of contributions per country, but at the end we discard this option since it was not entirely clear in many leaderboard entries. In general, we have been utterly cautious in dealing with informal data sources, triangulating the data as much as possible and discarding information not reliable enough. It is worth to mention that the type of analysis that we perform does require access to non-peer-reviewed sources, since the type of information sought appears in scientific literature only scarcely. 

\subsection{Data Annotation}

We apply content analysis~\cite{kovrigin2024importancereasoningcontextretrieval} to the collected data in order to elicit coding schemas related to several attributes. We apply a combination of deductive and inductive coding: we predefine the list of themes (corresponding to the different attributes we want to analyse) and, for each of them, we start with an empty set of codes that we gradually refine as we process the leaderboards' contents. We present below the result.

\subsubsection{Submitter Category}
\label{sec:method:inspection:submitterCategory}

To answer RQ2, we collect the type of organization that submitted a solution to \swebench{}.
Our inductive coding approach results in the following codes grouped into categories:

\begin{enumerate}[leftmargin=*, labelsep=1em]
\item \textbf{Academia}: The submission is made by members affiliated with academic institutions and is accompanied by a research article whose authors are all from academia. We distinguish among \textbf{Single Academy} institutions and \textbf{Collaboration Academia}, when different academic institutions collaborate in a submission.

\item \textbf{Industry}: The submission is made by members affiliated with a company. We further classify these companies by size, using the Microsoft-LinkedIn classification system\footnote{Microsoft Company Size Codes: \url{https://learn.microsoft.com/en-us/linkedin/shared/references/reference-tables/company-size-codes}}, which estimates the company size based on LinkedIn user data: 
\begin{enumerate*}[label=(\alph*)]
\item \textbf{Small}: Companies with fewer than 50 employees. 
\item \textbf{Medium}: Companies with fewer than 500 employees. 
\item \textbf{Large}: Private companies (not publicly traded) with more than 500 employees. 
\item \textbf{Large-Publicly Traded}: Publicly traded companies with more than 500 employees. 
\item \textbf{Unknown}: Companies for which we were unable to subclassify due to lack of information.
\end{enumerate*}
In addition, similar to the former case, we record \textbf{Collaboration Industry} when the submission comes from a joint effort between industry-affiliated contributors.

\item \textbf{Academia-Industry:} The submitter has a blend of academia and industry. We find two different types:
\begin{enumerate*}[label=(\alph*)]
\item \textbf{Collaboration Academia-Industry}: collaboration among academic and industry-affiliated contributors, as evidenced by an article co-authored by individuals from both domains.
\item \textbf{Academic Spin-off}: A company that is a spin-off from an academic institution, as explicitly indicated on its website, including reference to the originating university.
\end{enumerate*}

\item \textbf{Open-Source Community}: The submission is made by a non-profit or community-driven organization.

\item \textbf{Single Developer}: The submission is made by an individual acting in a personal capacity.

\item \textbf{Unknown}: We have not found enough information as to classify the submitter.
\end{enumerate}

\subsubsection{Product and Accessibility} 

To answer RQ3, we analyze how solutions submitted to \swebench{} are made accessible to users or stakeholders, focusing on two key dimensions.
First, we assess the \textbf{type of product}, if any, associated with each submission. 
In particular, we focus on two dimensions.
\begin{itemize}[leftmargin=*, labelsep=1em]
    \item {\bf Product Purpose}: describes \emph{what} the product does.
We identify twelve different purposes:
\begin{enumerate*}[label=(\alph*)]
\item Agent Framework,
\item Coding Assistant,
\item Development Assistant,
\item Development Framework,
\item Development Platform,
\item Issue Resolution,
\item Other-Inference Service,
\item Platform Code Optimization/Improvement,
\item Platform Code Representation,
\item Platform Integration, and
\item Problem Solving.
\end{enumerate*}
An additional code is introduced for products for which we cannot identify the form.

\item {\bf Product Form}: describes \emph{how} the product is delivered or used.
We identify five different forms:
\begin{enumerate*}[label=(\alph*)]
    \item Cloud platform,
 \item Command-line tool,
 \item Github plugin,
 \item IDE plugin, and
 \item Library.
\end{enumerate*}
An additional code is introduced for entries without form.
\end{itemize}

For example, the entries corresponding to the \texttt{Amazon Q Developer Agent} approach -submitted to both \shortlite{} and \shortverif{}- are linked to a Coding Assistant, distributed as an IDE plugin.\footnote{Amazon Q plugin: \url{https://plugins.jetbrains.com/plugin/24267-amazon-q/?b=jb&p=build&s=hero}}

Second, we record the \textbf{availability of the product}. Our inductive coding on both leaderboards results in four codes representing different levels of availability: 
\begin{enumerate}[leftmargin=*, labelsep=1em]
\item \textbf{Publicly Available Product (PAP)}:
A product based on the submitted solution is publicly accessible (e.g., DE plug-in available in plug-in/extensions marketplaces), either commercially (paid) or for free. 
Importantly, this category does not refer to the availability of the product' source code. 
\item \textbf{Upon Request (UR)}:
The product is not openly accessible but can be obtained upon request or through a waiting list. This includes early-access solutions and invite-only services.
\item \textbf{Business-to-Business (B2B)}:
The solution is only available under business agreements (B2B) and is not accessible to the general public. 
\item \textbf{Non-Commercial Solution (NCS)}: 
In this case, no associated product is distributed through plugin marketplaces nor is there a deployed solution such as an online IDE. 
This category typically includes tools shared as research artifacts or open-source code (e.g., on GitHub).
\item \textbf{Unavailable (UN)}: 
The solution is not available under any of the previously defined levels.
\end{enumerate}
Note that in this analysis, we focus on products associated to each entry -if any- rather than on the code used in the \swebench{} experiment (which could be eventually the same).

\subsubsection{Open-source solution}

To answer RQ4, we distinguish entries whose source code is open and available on platforms such as GitHub, from those whose code is closed. 
This assessment does not take into account the openness of the underlying language models, which is analyzed in RQ5.

\subsubsection{LLMs Employed in Solutions}

To answer RQ5, we identify the underlying LLMs used by each solution, noting that multiple models may be involved.
This information is usually presented in the submission name (shown in the leaderboard) but also eventually in the \texttt{metadata.yaml} file required for the \swebench{} submission.
If it is not available there, we search for this information in other sources such as preprint, scientific articles or blog posts.\footnote{If an article presents evaluations of multiple models, we select the one whose reported precision (i.e., \% Resolved) matches that of the corresponding leaderboard entry.}
Note that these articles may also report additional results (e.g., using other models) that are not included in the leaderboard. However, for this research question, we exclusively consider the result corresponding to the leaderboard submission, disregarding the others.

\section{Results}
\label{sec:results}

\begin{figure*}[t]
\begin{subfigure}[h]{0.49\textwidth}
  \centering
\includegraphics[width=\textwidth]{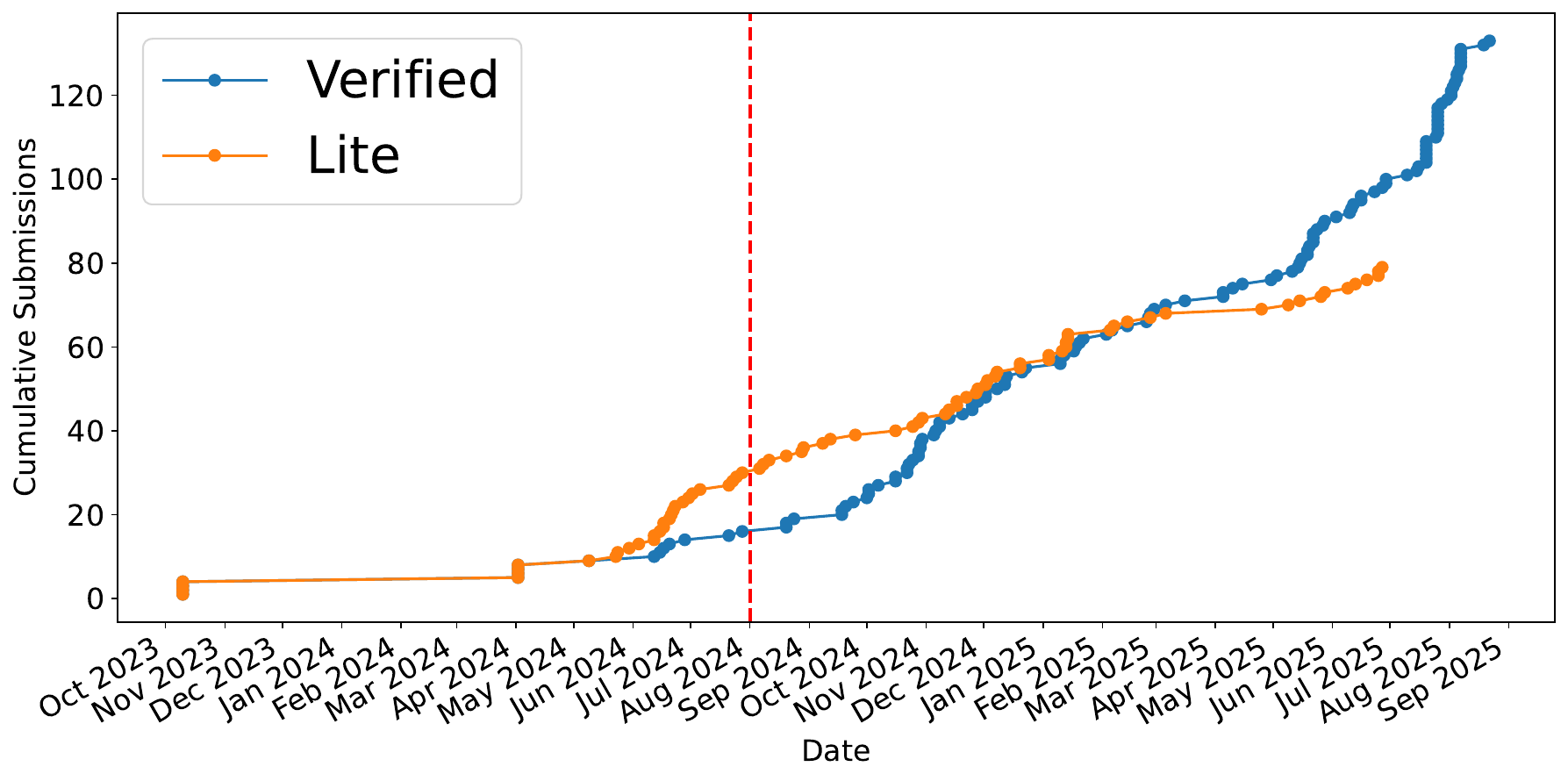}
\caption{Entries by date. The vertical dotted line correspond to the official release of \sweverif{} by OpenAI.}
\label{fig:date_submission}
\end{subfigure}
~
\begin{subfigure}[h]{0.49\textwidth}
  \centering
\includegraphics[width=\textwidth]{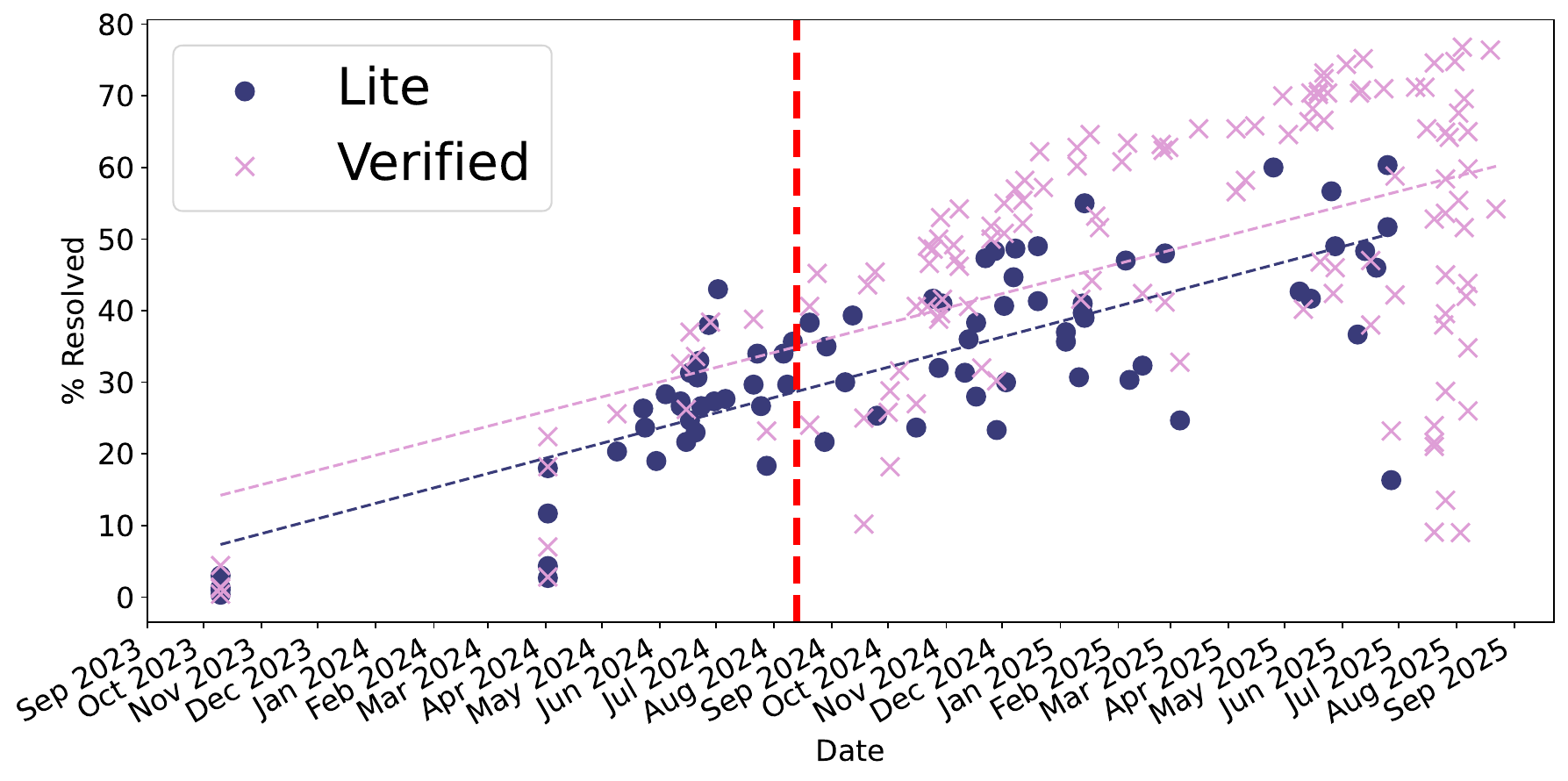}
\caption{Evolution of \precision{}. The vertical line correspond to the official release of \sweverif{} by OpenAI.}
\label{fig:resolved_evolve}
\end{subfigure}
\caption{}
\end{figure*}

\subsection{RQ1: Evolution of submissions and precision}

By September 18th 2025, we count, in total, 79 and 133 entries on \shortlite{} and \shortverif{}, respectively. 
We identify 52 and 59 distinct approaches, respectively. 
Among these, 29 approaches appear only in \shortlite{}, 36 only in \shortverif{}, and 23 in both.
The complete list of approaches and submissions by leaderboard is provided in the appendix~\cite{Apprendix}.

Fig. \ref{fig:date_submission} shows the cumulative number of entries on each leaderboard, taking the dates from the leaderboards themselves. 
The vertical dotted red line shows the release date of \sweverif{} (August 13th 2024), considered to be the date in which OpenAI released the blog article that describes the leaderboard\footnote{\url{https://openai.com/index/introducing-swe-bench-verified/}}. 
We observe that there are entries whose date precedes that official release, being the reason that the mentioned article also included, in addition to well defined construction criteria of \shortverif{}, the results of an evaluation of open-source approaches (\emph{open-source scaffolds} according to that article) such as AutoCoderRover~\cite{Zhang2024AutoCoderRover}, Agentless \cite{xia2024agentlessdemystifyingllmbasedsoftware}  and SWE-Agent \cite{yang2024sweagentagent} on \shortverif{}, which were already present in the \shortlite{} leaderboard.
In \shortverif{} leaderboard, these entries are listed with the same submission dates as in \shortlite{}.\footnote{Added in the leaderboard by this commit: \url{https://github.com/SWE-bench/swe-bench.github.io/commit/8d66692d5f55101c3c0f53daac3f8a479d29406e}}
The figure shows how the leaderboards have evolved along five stages:
\begin{enumerate}[leftmargin=*, labelsep=1em]
    \item Non-existence of \shortverif{}. From Oct. 2023 until May 2024, all submission were in \shortlite{} only.
    \item Prevalence of \shortlite{}. Until mid-September 2024, both repositories co-existed but still the growth rate of \shortlite{} was greater than that of \shortverif{}.
    \item Prevalence of \shortverif{}. In only two months, until mid-Nov. 2024, \shortverif{} experienced a sudden growth until reaching the same size as \shortlite{} (41 entries each).
    \item Similarity of both leaderboards. Until mid-Feb. 2025, the number of entries remained approximately the same for both \shortlite{} and \shortverif{}.
    \item Reduced activity on \shortlite{}. While submissions to \shortverif{} continue to grow steadily, activity on \shortlite{} has significantly dropped down during 2025.
    In fact, as of September 18th, the most recent entry listed in \shortlite{} is dated June 27th, 2025.
\end{enumerate}

Figure~\ref{fig:resolved_evolve} shows the evolution of the \precision{} metric, which is the primary measure reported in both leaderboards. 
Despite a period among Oct. 2024 and Nov. 2024 in which some solutions uploaded to \shortverif{} were not optimal, the precision is notably higher in \shortverif{} than in \shortlite{}.  Specifically, the median and maximum precision are 46.6\% and 76.8\% for \shortverif{}, compared to 32\% and 60.3\% for \shortlite{}.
This difference may be attributed to the construction of \shortverif{}, which was derived from the original \swebench{} by manually filtering out instances considered too hard or unsolvable~\cite{chowdhury2024swebenchverified}. 
We also observed in \shortverif{} several recent submissions (August 2025) with non-optimal results; most correspond to a single approach (\approach{mini-SWE-Agent})~\cite{yang2024sweagentagent} tested with different LLMs. 

\begin{shaded}
\underline{\bf{Answer to RQ1 (Submissions and precision):}}
Submissions have steadily increased over time, with recent activity concentrated on \shortverif{}, where the highest reported precision reaches 76.8\%, significantly overcoming \shortlite{} results and positioning well in the APR state of the art. 
\end{shaded}

\setlength{\tabcolsep}{3pt} 
\begin{table}[t!]
\centering
\scriptsize

\begin{tabular}{|l|llrll|llrll|rr|}
\hline
& \multicolumn{5}{|c|}{\swelite{}} & \multicolumn{5}{|c|}{\sweverif{}} & \multicolumn{2}{|c|}{Total} \\ 
\cline{2-11}
Submitter Type
& \multicolumn{3}{|c|}{} & \multicolumn{2}{|c|}{\precision{}}
& \multicolumn{3}{|c|}{} & \multicolumn{2}{|c|}{\precision{}} & \multicolumn{2}{|c|}{} \\
\cline{2-11}
& \#E & \#S & \%OS & Med & Max
& \#E & \#S & \%OS & Med & Max
& \#E & \#S \\
\hline
Single Academy (SA)         & 19 & 4 & \cca{95} & \ccb{23}   & \ccg{56.6} & 39 & 3 & \cca{97} & \ccb{38}   & \ccg{67.6} & 58 & 4 \\
Collab. Academia (CA)       &  6 & 3 & \cca{100}& \ccb{41.3} & \ccg{60.3} &  3 & 2 & \cca{100}& \ccb{32.8} & \ccg{46}   &  9 & 4 \\
Total Academia              & 25 & 7 &   --    & \ccb{27.3} & \ccg{60.3} & 42 & 5 &   --    & \ccb{36.4} & \ccg{67.6} & 67 & 8 \\
\hline
Academia Spinoff (AS)       &  2 & 1 & \cca{100}& \ccb{24.8} & \ccg{30.6} &  4 & 2 & \cca{75}  & \ccb{42.3} & \ccg{51.6} &  6 & 2 \\
Collab. Ac-Ind (CAI)  & 10 & 9 & \cca{80} & \ccb{29.8} & \ccg{47}   &  9 & 8 & \cca{78} & \ccb{53}   & \ccg{64.6} & 19 & 14 \\
Total Academia-Industry     & 12 &10 &   --    & \ccb{29.8} & \ccg{47}   & 13 &10 &   --    & \ccb{46.2} & \ccg{64.6} & 25 & 16 \\
\hline
Company-Small (CS)          & 19 &15 & \cca{47} & \ccb{38}   & \ccg{60}   & 33 &17 & \cca{45} & \ccb{53.2} & \ccg{74.4} & 52 & 24 \\
Company-Medium (CM)         &  1 & 1 & 0  & \ccb{49}   & \ccg{49}   &  9 & 5 & \cca{22} & \ccb{64.6} & \ccg{73.2} & 10 & 5 \\
Company-Large (CL)          &  4 & 2 & 0  & \ccb{34.8} & \ccg{39.3} &  5 & 2 & \cca{40} & \ccb{70.6} & \ccg{75.2} &  9 & 3 \\
Comp.Pub. Traded (CLP)      &  7 & 5 & \cca{14} & \ccb{27.3} & \ccg{48.3} & 25 & 8 & \cca{32} & \ccb{42.4} & \ccg{76.8} & 32 & 10 \\
Collab.Industry (CI) &  --& --&   --    &    -       &    -       &  1 & 1 & \cca{100}& \ccb{46.8} & \ccg{46.8} &  1 & 1 \\
Company-Unknown (CU)        &  1 & 1 & \cca{100}& \ccb{44.6} & \ccg{44.6} &  1 & 1 & 0  & \ccb{71.2} & \ccg{71.2} &  2 & 2 \\
Total Industry              & 32 &24 &   --    & \ccb{35.8} & \ccg{60}   & 74 &34 &   --    & \ccb{53.7} & \ccg{76.8} & 106 & 45 \\
\hline
OS Community (OSC)          &  6 & 2 & \cca{100}& \ccb{28.69}& \ccg{39}   &  2 & 2 & \cca{100}& \ccb{67.1} & \ccg{70.8} &  8 & 3 \\
Single Developer (SD)       &  2 & 2 & \cca{50} & \ccb{29}   & \ccg{30.3} &  --& --&   --    &    -       &    -       &  2 & 2 \\
\hline
Unknown (UNK)               &  2 & 2 & \cca{50} & \ccb{42.2} & \ccg{42.6} &  2 & 2 & \cca{50} & \ccb{58.9} & \ccg{74.6} &  4 & 3 \\
\hline
\end{tabular}
\caption{Descriptive statistics of submitter organizations. \#E shows the total number of entries, \#S is the number of distinct submitters. \%OS is the share of open-source involvement.}
\label{tab:summarySubmitter}
\end{table}

\subsection{RQ2: Types of Submitters}

\label{res:rq1:submitter}

\label{sec:results:submitter}

Table~\ref{tab:summarySubmitter} summarizes, for each submitter type, the number of entries (\#E), the number of distinct submitters (\#S), and the median and maximum \precision{} scores. 
%
We observe that the majority of submitters are companies, representing 58\% of the total; this share increases to 79\% when industry–academia collaborations are included.
The proportion of industrial submitters is larger in \shortverif{} (64\%) than in \shortlite{} (51\%).

In terms of leaderboard entries, the difference in the number of entries between \shortlite{} (79) and \shortverif{} (133) comes mainly from a larger number of industry submissions, increasing from 41\% (32 entries) to 56\% (74 entries). 

Notable is the increase in submissions from large publicly traded companies, which account for 18.8\% (25 entries) of all entries in \shortverif{}, compared to  8.8\% (7 entries) in \shortlite{}.
The absolute number of academic submissions remains on a similar level (32\%), and open source community solutions virtually disappear in \shortverif{}, the same as single developer solutions. 
Solutions from small companies prevail in both leaderboards with 19 entries made by 15 distinct companies in Lite and 33 entries by 17 companies in Verified. 
We now examine the characteristics of each leaderboard individually.

\textbf{The case of \swelite{}}. The dominance of companies between the submitters is not as high as in \shortverif{} but still more than half (51\%), and from them, small companies prevail (up to 15 submitters, representing 32.6\% over the total). 
We also found two entries submitted by an Academia spin-off, AutoCodeRover, which has been acquired by an established company, Sonar, in Feb. 2025.\footnote{\url{https://www.sonarsource.com/company/press-releases/sonar-acquires-autocoderover-to-supercharge-developers-with-ai-agents/} (Last access Sept. 25, 2025)}

Large companies—including publicly traded and medium-sized firms—have also submitted their solutions, typically independently and without external collaboration (e.g., IBM, Amazon).
In contrast, some companies submitted exclusively through collaborative efforts, primarily with academic partners: Google \cite{su2025learnbyinteract}, Intel \cite{yu2025orcaLocallmAgent} and Meta \cite{wei2025SWE-RL}. Other collaborations have involved small companies founded in recent years such as All-Hands \cite{wang2024openhandsopenplatformai}.

This shows that the interest in \swebench{} is not limited to academia—where it originally emerged—or academic spin-offs, but also extends to established players in the software industry.
This growing diversity of contributors suggests that \swebench{} is becoming a widely accepted benchmark for evaluating AI-based issue repair tools across sectors. 

Seven submitters to \shortlite{} are from academia, one of them being Princeton University, the founder of \swebench{} \cite{jimenez2024SWEBenchLLMs}. 
We also found entries submitted by two open source communities (\approach{Moatless} and \approach{Aider}).
Lastly, two submissions were made by individual developers: \approach{SIMA} which tested the multi-agent architectures proposed by Li et al.~\cite{li2024MoreAgentsneed}, and \approach{Aegis}, developed as a side project.

\textbf{The case of \sweverif{}}. As commented, the prevalence of industry in \shortverif{} increases significantly: 
the 64\% of the submitters and the 56\% of the entries are from industry.
Per type of companies, fluctuations are minor; we may remark, though, the involvement of NVIDIA, Google, Anthropic and EPAM. Concerning academia, a new collaboration between two U.S. universities (UCSB and Columbia) contributed to \shortverif{}.

\paragraph{Analysis of \% Resolver rate}

Table~\ref{tab:summarySubmitter} shows the median and maximum \precision{} achieved by each type of submitter. 

In \swelite{}, the highest precision (60.3\%) is achieved by \approach{ExpeRepair}~\cite{mu2025experepairdualmemoryenhancedllmbased}, a submission resulting from a collaboration between academic institutions. A close second (60\%) is the submission by a small company, \approach{Refact.ai}.
Notably, recent entries from individual academic groups also surpass the 50\% mark, including \approach{SWE-Agent} \cite{yang2024sweagentagent} (56.67\%) and \approach{SemAgent}~\cite{pabba2025semagentsemanticsawareprogram} (51\%).
Surprisingly, no medium or large company has outperformed these results so far.
Across all types of submitters, the median precision remains relatively consistent, ranging from 27\% to 29.8\%, with the exception of industry submissions, which exhibit a notably higher median (35.8\%).
We assess whether there are statistically significant differences in precision across submitter types using the Kruskal–Wallis test. 
The analysis reveals a statistically significant difference between Single Academia and Small Companies (H = 19.89, p = 0.0469).
This difference may be influenced by the early submissions to the leaderboard, which were primarily made by academic groups and tended to have lower precision.

In \sweverif{}, the results show a different landscape. 
Various types of companies —small, medium, and large— submitted results with the highest \precision{}, all exceeding 73\%.
However, other submitter categories also achieved competitive results. Notably, academia (e.g., Princeton University with \approach{SWE-Agent}) and the open-source community (e.g., \approach{Moatless}) reached precision scores above 66\%.
Excluding the open-source submissions —only two entries, considered outliers—the highest median precision is observed among medium and large companies, driven by submissions from Anthropic, which evaluated its state-of-the-art LLMs featuring hybrid reasoning capabilities.
In contrast, the median precision from academia is notably lower than that of the other categories.
The Kruskal–Wallis test returned a similar result as \shortlite{} (H = 38.0953, p = 0.0001). 
Dunn’s post-hoc test further reveals that the precision scores from small, medium and large companies differ significantly from those of academia ($p\_values$ 0.01475, 0.00527 and 0.01366, resp.). 
This finding corroborates the trend discussed above.

\begin{figure*}[t!]
    \centering
    \begin{subfigure}[t]{0.5\textwidth}
        \centering
        \includegraphics[width=\textwidth]{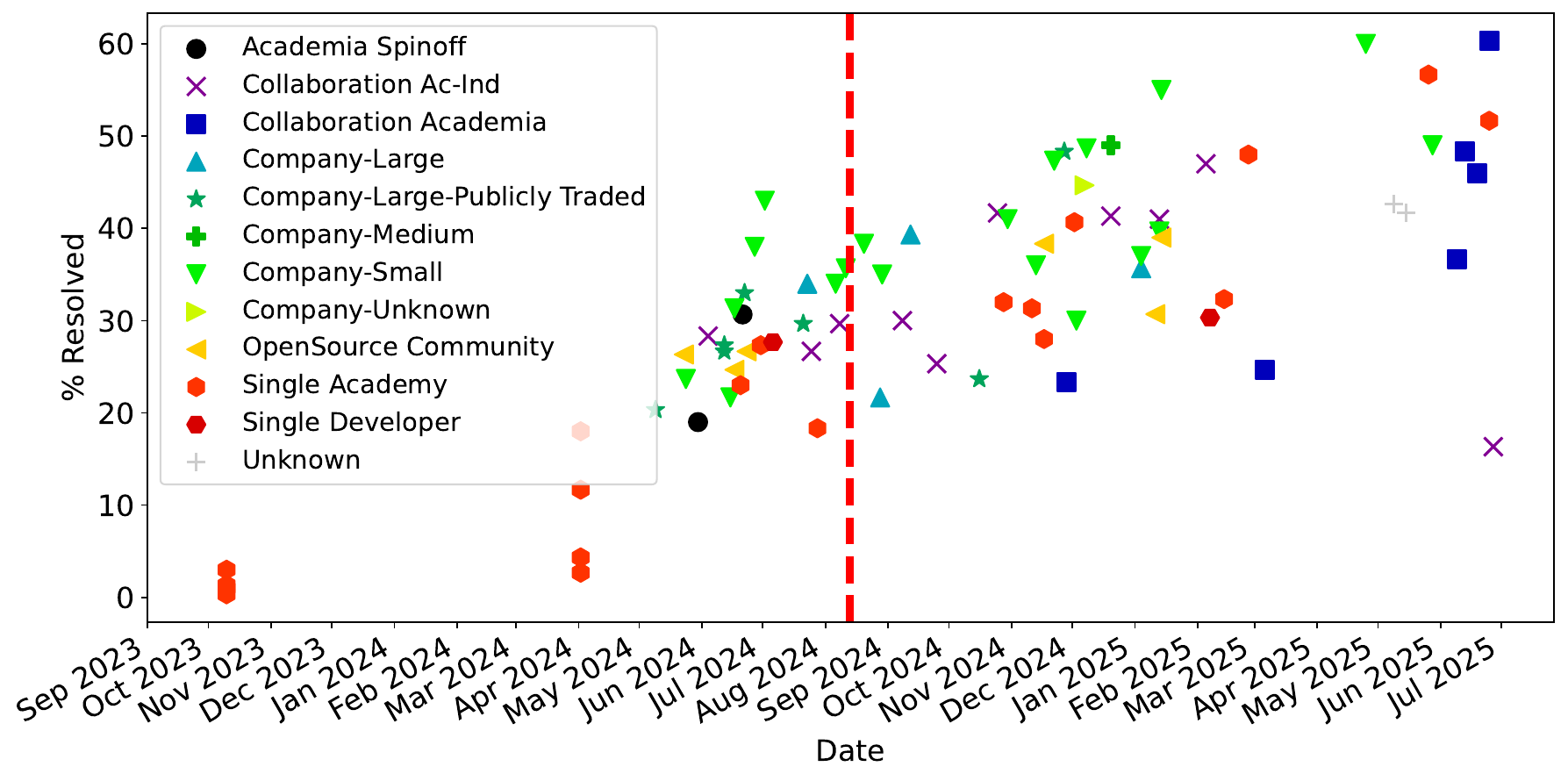}
        \caption{\swelite{}}
        \label{fig:evolLiteByOrigin}
    \end{subfigure}%
     ~ 
      \begin{subfigure}[t]{0.5\textwidth}
        \centering
        \includegraphics[width=\textwidth]{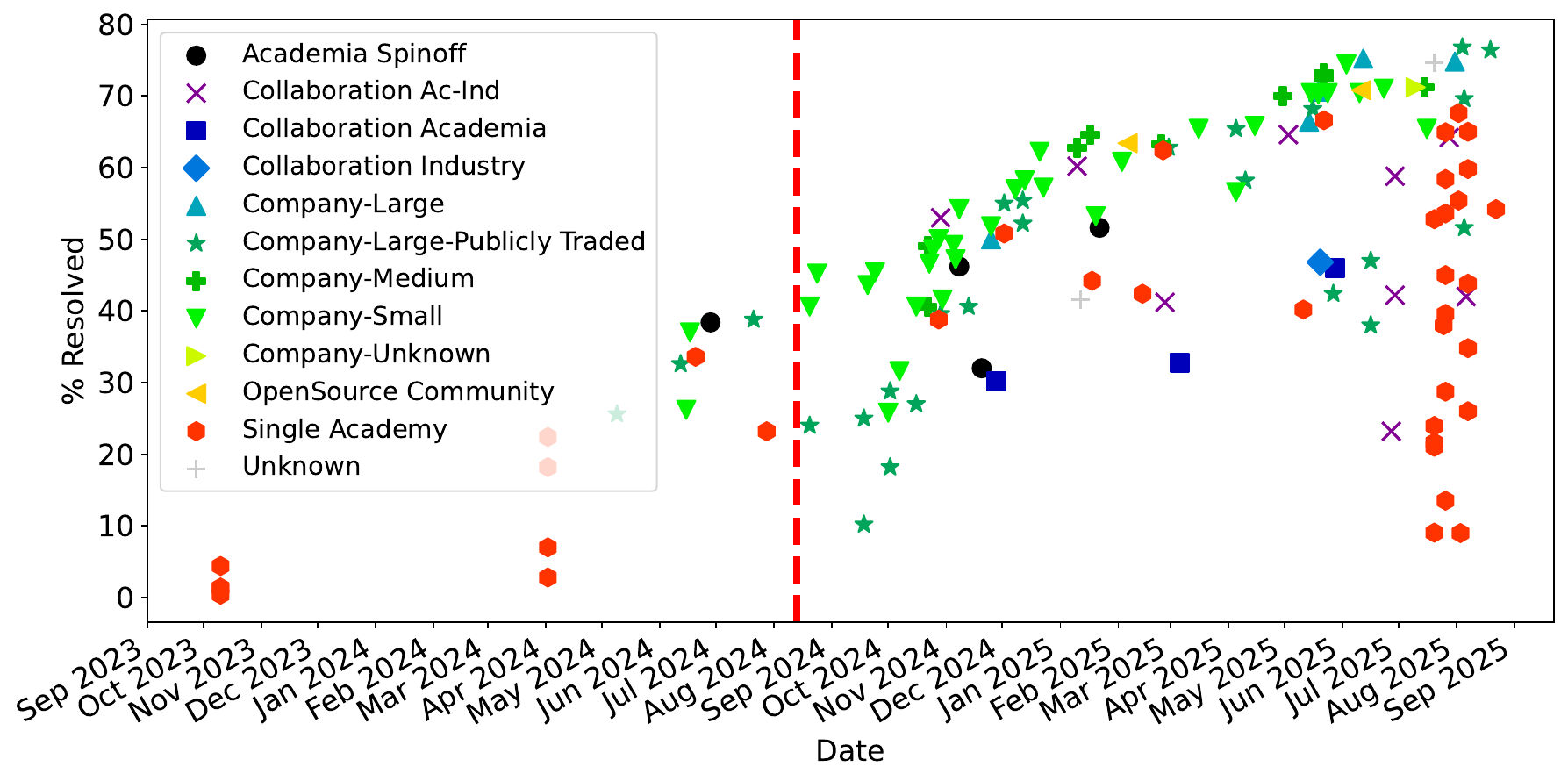}
        \caption{\sweverif{}}
        \label{fig:evolVerifByOrigin}
    \end{subfigure}%
    \caption{Evolution of \texttt{\% Resolved} by type of submitter.}
    \label{fig:evolPrecisionByOrigin}
\end{figure*}


\paragraph{Temporal Evolution of \precision{}}

Fig~\ref{fig:evolPrecisionByOrigin} shows the leaderboards' entries arranged according to the date and precision displayed in the leaderboards themselves.
Firstly, we observe that the initial entries from both leaderboards come from academia. 
These early entries achieved a maximum precision of around 20\%. 
Then, starting in May–June 2024, both leaderboards began receiving submissions from other organizations beyond academia.

In \shortlite{}, as shown in  Figure~\ref{fig:evolLiteByOrigin} we observe that small companies have been constantly pushing the results. 
Notably, \approach{Isoform} achieved a state-of-the-art precision of 55\% on Jan. 2025, and a months later Refact.ai achieved a 60\%.
More recently academic submissions have achieved competitive results, led by a collaboration between institutions (\approach{ExpeRepair})~\cite{mu2025experepairdualmemoryenhancedllmbased}, and \approach{SWE-Agent}, from a single academic team~\cite{yang2024sweagentagent}.
In \shortverif{}, as shown in Fig.~\ref{fig:evolVerifByOrigin}, small companies have also consistently driven improvements in precision. For instance, \texttt{Refact.ai} achieved top precision (74.40\%) on June 2025.
Nevertheless, other types of companies reached state-of-the-art performance as well —for example, the medium-sized AI leader \texttt{Anthropic} achieved 73.20\%, the large company \texttt{Bytedance} achieved a 75.2\% with \approach{TRAE}~\cite{traeresearchteam2025traeagentllmbasedagent}.
and the publicly traded company EPAM currently holds the top entry with 75.2\% with its \approach{AI/Run Developer Agent}.
Submissions from academia have achieved competitive, though not top-ranking, results -driven by \texttt{Agentless-1.5} \cite{xia2024agentlessdemystifyingllmbasedsoftware} (50.8\%) and \texttt{SWE-Agent} \cite{yang2024sweagentagent} (66\%), both developed by single academic teams, as well as by a collaboration between the University of California (UCSB) and Meta, which produced \texttt{PatchPilot}~\cite{li2025patchpilot} with near state-of-the-art precision of 64.6\% on May 2025.

\begin{shaded}
\underline{\bf{Answer to RQ 2 (Submitters):}}
Entries in the leaderboards originate from a wide variety of submitters, including (combinations or particularization of) academy, industry (dominated by 24 different companies submitting 52 entries) and individual developers. 
This diversity underscores the benchmark’s accessibility and minimal barriers to participation, demonstrating that it is readily approachable for experimentation by a broad spectrum of contributors.
Since the introduction of \shortverif{}, companies (especially small ones) have been driving progress toward state-of-the-art performance. From these results, a recommendation to research centers is to seek collaboration with industry actors in order to deliver better solutions. 
\end{shaded}


\subsection{RQ 3: Products and
Availability Modes}

\label{sec:results:product}

\newcolumntype{b}{X}
\newcolumntype{s}{>{\hsize=.05\hsize}X}
\newcolumntype{m}{>{\hsize=.5\hsize}X}

\setlength{\arrayrulewidth}{0.3pt} 

{
\setlength{\tabcolsep}{3pt} 
\begin{table}[t!]
\centering
\scriptsize
\begin{tabular}{|l|ll|c|c|cc|}
\hline
& \multicolumn{2}{c}{Product} 
& \multicolumn{1}{|c|}{Lite} 
& \multicolumn{1}{|c|}{Verified} 
& \multicolumn{2}{|c|}{Total} \\
\cline{2-7}
\rotatebox{0}{Av.} & Purpose & Form 
& \% & \% & \#A & \#E \\
\hline

\multirow[t]{12}{*}{PAP} 
& Coding Assistant & IDE plugin & \ccb{33.5} & \ccb{44.2} & 10 & 27 \\ 
& \multirow[t]{4}{*}{Development Assistant} 
  & Cloud platform & \ccb{37} & - & 1 & 1 \\
&  & Command-line tool & - & \ccb{64.6} & 1 & 1 \\
&  & Github plugin - Bot & \ccb{40.5} & \ccb{62.2} & 3 & 3 \\
&  & lug-in, Cloud & - & \ccb{56.6} & 1 & 1 \\
& \multirow[t]{5}{*}{Development Platform} 
  & Cloud platform & \ccb{36.2} & \ccb{52.4} & 8 & 26 \\
&  & IDE & - & \ccb{71} & 1 & 1 \\
&  & Command-line, Plug-in, Cloud & \ccb{60} & \ccb{72.4} & 1 & 3 \\
&  & Plug-in, Cloud & - & \ccb{71.2} & 3 & 3 \\
&  & Plug-in, IDE, Terminal & - & \ccb{72.9} & 1 & 2 \\
& Other-Inference Service & Cloud platform & - & \ccb{40.6} & 1 & 1 \\
& Problem Solving & Foundation Model & - & \ccb{63.2} & 3 & 7 \\
& \multicolumn{2}{c|}{Total} & \ccb{36.5} & \ccb{55.8} & 34 & 76 \\
\hline

\multirow[t]{5}{*}{NCS} 
& Agent Framework & Command-line tool & \ccb{30.7} & \ccb{70.8} & 1 & 6 \\
& Coding Assistant & Command-line tool & \ccb{26.3} & - & 1 & 1 \\
& Development Assistant & Command-line tool & \ccb{38.3} & \ccb{74.6} & 2 & 3 \\
& Development Framework & Library & - & \ccb{63.4} & 1 & 1 \\
& Issue Resolution & Command-line tool & \ccb{27.3} & \ccb{38.4} & 19 & 75 \\
& \multicolumn{2}{c|}{Total} & \ccb{28.3} & \ccb{39.6} & 24 & 86 \\
\hline

\multirow[t]{3}{*}{B2B} 
& \multirow[t]{2}{*}{Development Platform} 
  & Command-line tool & - & \ccb{47.5} & 1 & 6 \\
&  & Command-line, Plug-in, Cloud & \ccb{48.3} & - & 1 & 1 \\
& Code Opt./Impr. & Unknown & \ccb{24.8} & \ccb{46.2} & 2 & 5 \\
& \multicolumn{2}{c|}{Total} & \ccb{30.7} & \ccb{46.2} & 4 & 12 \\
\hline

\multirow[t]{5}{*}{UR} 
& Code Migration & None & - & \ccb{71.2} & 1 & 1 \\
& Development Platform & Cloud platform & \ccb{30} & \ccb{52.2} & 2 & 2 \\
& Code Opt./Impr. & Cloud platform & - & \ccb{32} & 1 & 1 \\
& Platform Code Representation & None & - & \ccb{53.2} & 1 & 1 \\
& Platform Integration & None & \ccb{45} & - & 1 & 2 \\
& \multicolumn{2}{c|}{Total} & \ccb{35} & \ccb{52.7} & 6 & 7 \\
\hline

\multicolumn{3}{|c|}{UN} 
& \ccb{35.7} & \ccb{47} & 20 & 31 \\
\hline

\end{tabular}
\caption{Product Availability (column Av.), Product Purpose and Form. 
\#A is the number of distinct approaches, \#E is the number of leaderboard's entries, \% is the median precision. 
The availabilities are PAP: Publicly Available Product, UR: Upon Request, B2B: Business-to-Business, NCS: Non-Commercial Solution, UN: Unavailable. 
Full table, broken down by leaderboard, in the appendix.}
\label{tab:resultsProductTypeAndAvailability}
\end{table}
}

Table~\ref{tab:resultsProductTypeAndAvailability} presents the number of approaches (\#A) and entries (\#E) per product availability category, along with their median precision (\precision{}). We analyse below the contents of each category.



{\bf Publicly Available Product (PAP).} It includes publicly accessible commercial tools and accounts for 76 entries across both leaderboards, representing 34 distinct approaches.
The submissions related PAP products achieve the highest median \precision{} across both leaderboards (36.5\% and 55.8\%). 
The most common product type from this category is \emph{Coding Assistant}, comprising 10 products available as IDE plug-ins and linked 27 entries. 
These include, for example, \texttt{Amazon Q Developer} and \texttt{Alibaba Lingma}, available as extensions for popular IDEs. 
The second major product type is \emph{Development Platform}, typically delivered as cloud-based services (26 entries).
These platforms, such as \texttt{OpenHands} and \texttt{Solver}, generally allow developers to specify new features using natural language and integrate directly with code repository platforms (e.g., GitHub).
Other solutions such as \texttt{Refact.ai} offer different product forms in addition to the cloud Services, such as IDE plugins.
%
A third type includes \emph{Development Assistant}, which helps developers in particular development task. Three of the entries, such as \texttt{CodeStory Aide},  materialize this assistant as automated bots integrated into GitHub workflows, interacting with developers via pull requests, while the two others as a command-line tool and a cloud platform. 
%


The {\bf Non-Commercial Solutions (NCS)} category accounts for 86 entries across both leaderboards, corresponding to 24 distinct approaches.
The dominant product type is \emph{Issue Resolution}, comprising 19 approaches such as Agentless \cite{xia2024agentlessdemystifyingllmbasedsoftware}, SWE-agent \cite{yang2024sweagentagent}, and HyperAgent \cite{phan2024hyperagent}.
This category also includes extensions of these tools, for instance, OrcaLoca~\cite{yu2025orcaLocallmAgent}. 
We also identify a \emph{Coding Assistant}, namely \texttt{Aider}, which differs from the previously mentioned assistants in that it is implemented as a command-line tool rather than as a plug-in.


The {\bf Business-to-Business (B2B)} category includes 4 distinct approaches and 12 entries.
It covers two main product types: \emph{Development Platform} and \emph{Platform Code Optimization}.
These tools are typically integrated into enterprise workflows and tailored for internal use.


The {\bf Under Request (UR)} category is associated with 7 entries from 6 approaches across both leaderboards.
It comprises a variety of product types, including \emph{Platform Integration}, \emph{Code Representation}, \emph{Code Optimization}, and even a \emph{Development Plugin}.
These products are typically available only upon request and are represented by a small number of entries.
Moreover, submissions in this category are generally limited to a single leaderboard, rather than appearing in both.

Finally, we grouped under {\bf UN} (Unavailable) those entries for which no associated product could be identified.
This group comprises 20 approaches and 31 entries. 

Note that the products presented above are not necessarily the exact artifacts used to conduct the \swebench{} experiments.
In some cases, such as \texttt{Agentless} or \texttt{SWE-Agent}, the systems were specifically designed for that purpose, and the artifacts available are used for replicating the study. In contrast, some tools, such as general-purpose Coding Assistants, were not originally designed for repair tasks, and their use in \swebench{} likely reflects the initiative of submitters to adapt existing capabilities to meet the benchmark requirements.

\begin{shaded}
\underline{\bf{Answer to RQ 3 (Product):}}
Entries in the \swebench{} leaderboards are made available to the community in different ways: publicly accessible commercial tools, non-commercial solutions, available under request or as B2B solutions. Entries associated with Publicly Available Products and Under-Request solutions stand out for their precision, particularly in \shortverif{}, due to a larger involvement of industry actors.
\end{shaded}

\subsection{RQ 4: Open- and Closed- Source Solutions}
\label{sec:results:opensource}

\begin{table}[t!]
\small
\begin{tabular}{|l|rr|ll|rr|ll|}
\hline
&\multicolumn{4}{c|}{\swelite{}} & \multicolumn{4}{c|}{\sweverif{}} \\ 
\cline{2-9}
&\multicolumn{2}{c|}{} & \multicolumn{2}{|c|}{\precision{}}
&\multicolumn{2}{c|}{} & \multicolumn{2}{c|}{\precision{}}\\
\cline{2-9}
&\#E & \#S  & Med & Max
&\#E & \#S & Med & Max\\
\hline
Open-source & 53 & 27 & \ccb{30.33} & \ccg{60.33} & 82 & 28 & \ccb{46.1} & \ccg{75.2} \\
Closed-source & 26 & 20 & \ccb{35.67} & \ccg{55} & 51 & 27 & \ccb{52.2} & \ccg{76.8} \\

\hline
\end{tabular}
\caption{Open-source vs Closed-source solutions. 
\#E is the number of entries, \#S is the number of distinct submitters.}
\label{tab:results:opensource}

\end{table}

Table~\ref{tab:results:opensource} presents the number of entries associated with open-source and closed-source solutions.
In \shortlite{}{}, the 67\% of entries (53 in total, from 27 unique submitters) are open-source.
In \shortverif{} that percentage  is lower: 61.6 \% (82 entries from 28 distinct submitters).
Beyond this, the lower share of open-source entries can be explained by \shortverif{}{}’s higher proportion of industry submissions, particularly from startups and established companies that typically do not release their systems as open source. As shown in Table~\ref{tab:summarySubmitter} (column \%OS), entries associated with companies fall below 47\% open source. However, collaborations—either between companies (1 entry) or between industry and academia (19 entries)—raise this share to at least 78\%. The table also shows that whenever academia is involved (whether as the sole submitter, in collaboration, or through a spin-off), most of the resulting solutions are open source.
This underlines academia’s central role in driving openness within the \swebench{} ecosystem.

Regarding precision, closed-source entries generally achieve higher median scores on both leaderboards, yet open-source solutions remain highly competitive: the top performer in \shortlite{} is open source, and in \shortverif{} several open-source approaches achieve precision close to the leading closed-source entries.

\begin{shaded}
\underline{\bf{Answer to RQ 4 (Open \& Closed source):}}
As in many other domains, companies tend to deliver their solutions as closed-source, while academia drives openness. While closed-source solutions exhibit a higher median \precision{}, emerging open-source tools are showing competitive performance, with several of them achieving state-of-the-art results on both \swelite{} and \sweverif{} in 2025.
\end{shaded}

\subsection{RQ 5: LLMs Employed}
\label{sec:results:llms}

\begin{table}[t]
\scriptsize
\centering
{\setlength{\tabcolsep}{4pt}\renewcommand{\arraystretch}{0.95}%
\begin{tabular}{|l|r|rr|}

\hline
LLMs & \#E & \multicolumn{2}{c|}{\precision{}} \\
\cline{3-4}
     &     & Med & Max \\
    \hline
\multicolumn{4}{|c|}{\swelite{}} \\
\hline
Claude 4 Sonnet & 2  & \ccb{58.5}   & \ccg{60.33} \\
Claude 3.7 Sonnet+o4-mini & 1 & \ccb{60}     & \ccg{60} \\
No Info & 9 & \ccb{35.67} & \ccg{55} \\
Claude 3.7 Sonnet+Gemini 2.5 Pro & 1 & \ccb{51.67} & \ccg{51.67} \\
Claude 3.5 Haiku+Claude 3.5 Sonnet+Gemini 2.5 Pro & 1 & \ccb{49} & \ccg{49} \\
GPT-4 & 8 & \ccb{25.17} & \ccg{48.67} \\
Claude 3.5 Sonnet & 17 & \ccb{41} & \ccg{48.33} \\
Claude 3.5 Sonnet+o3-mini+o4-mini & 1 & \ccb{48.33} & \ccg{48.33} \\
Claude 3.7 Sonnet & 1 & \ccb{48} & \ccg{48} \\
Claude 3.5 Sonnet+DeepSeek R1 & 1 & \ccb{47} & \ccg{47} \\
Claude 3.5 Sonnet+GPT-4o & 4 & \ccb{35} & \ccg{43} \\
Claude 3.5 Haiku+Claude 3.5 Sonnet & 1 & \ccb{41.67} & \ccg{41.67} \\
GPT-4o & 14 & \ccb{28.835} & \ccg{39.33} \\
Deepseek V3 (*)& 2 & \ccb{33.685} & \ccg{36.67} \\
Claude 3.5 Sonnet+GPT-4 & 1 & \ccb{33} & \ccg{33} \\
o3-mini & 2 & \ccb{31.33} & \ccg{32.33} \\
Claude 3+GPT-4o & 2 & \ccb{25.83} & \ccg{26.33} \\
Qwen2.5 (*)& 3 & \ccb{23.33} & \ccg{24.67} \\
LLama 3+Mistral-Large+Qwen2.5+Granite (*)& 1 & \ccb{23.67} & \ccg{23.67} \\
GPT-4+GPT-4o & 1 & \ccb{21.67} & \ccg{21.67} \\
Claude 3 & 2 & \ccb{8} & \ccg{11.67} \\
Claude 2 & 1 & \ccb{3} & \ccg{3} \\
LLama 3 (*)& 2 & \ccb{1.165} & \ccg{1.33} \\
GPT3/3.5 & 1 & \ccb{0.33} & \ccg{0.33} \\
\hline
\multicolumn{4}{|c|}{\sweverif{}} \\
\hline
Claude 4 Sonnet & 10 & \ccb{71} & \ccg{76.8} \\
Claude 4 Sonnet+Claude 4 Opus+GPT-5+Gemini 2.5 Pro & 1 & \ccb{76.4} & \ccg{76.4} \\
Claude 3.7 Sonnet+Claude 4 Sonnet+Claude 4 Opus+Gemini 2.5 Pro & 1 & \ccb{75.2} & \ccg{75.2} \\
Claude 4 Sonnet+o4-mini & 1 & \ccb{74.4} & \ccg{74.4} \\
Claude 4 Opus & 2 & \ccb{70.4} & \ccg{73.2} \\
Claude 4 Sonnet+GPT-4o & 1 & \ccb{71.2} & \ccg{71.2} \\
Claude 3.7 Sonnet+Claude 4 Sonnet+GPT-4+Gemini 2.5 Pro & 1 & \ccb{71} & \ccg{71} \\
Claude 3.7 Sonnet+o1+o4-mini+Gemini 2.5 Pro & 1 & \ccb{70.6} & \ccg{70.6} \\
Claude 3.7 Sonnet+o3+o4-mini & 1 & \ccb{70.4} & \ccg{70.4} \\
Claude 3.7 Sonnet+o3+Gemini 2.5 Pro & 1 & \ccb{70.2} & \ccg{70.2} \\
Claude 3.7 Sonnet+o4-mini & 1 & \ccb{70} & \ccg{70} \\
Qwen3 (*)& 5 & \ccb{55.4} & \ccg{69.6} \\
Claude 3.7 Sonnet+o1+o3+o3-mini & 1 & \ccb{68.2} & \ccg{68.2} \\
Claude 3.7 Sonnet & 5 & \ccb{62.4} & \ccg{66.4} \\
No Info & 18 & \ccb{45.3} & \ccg{65.8} \\
Kimi-K2 (*)& 2 & \ccb{54.6} & \ccg{65.4} \\
Claude 3.7 Sonnet+o1 & 1 & \ccb{65.4} & \ccg{65.4} \\
GPT-5 & 2 & \ccb{62.4} & \ccg{65} \\
gpt-4o-mini & 2 & \ccb{58.9} & \ccg{64.6} \\
o1 & 1 & \ccb{64.6} & \ccg{64.6} \\
GLM-4.5 & 2 & \ccb{59.2} & \ccg{64.2} \\
Claude 3.5 Sonnet+Qwen2.5 & 1 & \ccb{63.4} & \ccg{63.4} \\
Claude 3.5 Sonnet & 15 & \ccb{51.6} & \ccg{62.8} \\
Claude 3.5 Sonnet+o3-mini & 1 & \ccb{60.8} & \ccg{60.8} \\
o3 & 1 & \ccb{58.4} & \ccg{58.4} \\
\hline

\end{tabular}}
\caption{Combinations of base LLMs. \#E is the number of entries; \precision{} shows median and maximum. An asterisk (*) indicates that all models are open-source. Top-25 by precision; full list in appendix.}
\label{tab:results:llmCombinationOneColumn}
\end{table}

Table~\ref{tab:results:llmCombinationOneColumn}  presents the base LLMs used in the submissions to both \shortlite{} and \shortverif{}, and how frequent these LLMs are used together in the submissions.
The table is sorted by \precision{} and shows the top-25.

The majority of submissions (55/79 in \shortlite{} and 94/133 in \shortverif{}) report using a single LLM. Excluding the most recent state-of-the-art models from the Claude 4 family, Table~\ref{tab:results:llmCombinationOneColumn} shows that top-performing results are often achieved through combinations of multiple LLMs. 
Notably, most LLM combinations appear in only one submission per leaderboard, with a few exceptions: \texttt{Claude 3.5 Sonnet + GPT-4o} is used by four submissions in \swelite{} and five in \sweverif{}.

In \shortlite{}{}, \llm{Claude 4 Sonnet} employed by the approach  \approach{ExpeRepair} \cite{mu2025experepairdualmemoryenhancedllmbased} achieved the highest performance with a preposition of 60.3\%.
Only two entries relied on this LLM alone (column \#E).
\llm{Claude 3.5 Sonnet} is the most frequently used LLM, appearing in 16 entries as a standalone model, and in several others in combination with additional LLMs.
For instance, four entries combine both \llm{Claude 3.5 Sonnet} and \llm{GPT-4o}.

In \shortverif{}, a similar trend is observed: 
the state-of-the-art result (76.8\%), by \approach{EPAM} (August 2025) was also achieved with \llm{Claude 4 Sonnet}.
In total, 10 entries just rely on this LLM.
Nevertheless, using combination of LLMs also led to competitive performance. 
For instance, \approach{TRAE} achieved 75.2\%
using 4 LLMs including \llm{Claude 4 Sonnet and Opus}, and \llm{Claude 3.7 Sonnet}.
\llm{Claude 3.5 Sonnet} remains the most frequently used model, appearing in 5 entries as a standalone LLM and in nine entries combined with others, such as \llm{Qwen2.5}.
The second most frequently used LLM in isolation is \llm{Qwen2.5}, with 12 entries. Notably, it is an open-source model; however, the solutions employing it achieved lower performance, with precision scores up to 47\%, which is why it does not appear among the top 25 in Table~\ref{tab:results:llmCombinationOneColumn}.

\paragraph{Openness of LLMs}
\label{sec:results:openessmodels}

The majority of submissions rely on proprietary, closed-source LLMs accessed via API, most notably from the Claude 3 and 4 model families. 
However, a range of open-source models has also been employed, typically after fine-tuning for specific tasks such as patch evaluation.
Combinations of fully open-source models are indicated in Table \ref{tab:results:llmCombinationOneColumn} with an asterisk (*).
The most frequently used open-source model is \llm{Qwen2.5}, serving as the sole model in 3 entries from \shortlite{} and 12 from \shortverif{}{}. 
Some entries rely exclusively on a set of open-source models, for example, one submission from IBM uses \llm{LLama}, \llm{Mistral-Large}, \llm{Qwen2.5} and \llm{Granite}.
In general, the precision of approach using only open-source LLMs is lower than those using closed LLMs, in both leaderboards.
Nevertheless, some solutions combine both closed- and open-source models. One such example is \texttt{AgentScope}, which achieved near state-of-the-art performance on \sweverif{} (63.4\%), at the time the result was published (February 2025), using \llm{Claude 3.5 Sonnet} in combination with \llm{Qwen2.5}.
Notably, no other submission using \llm{Claude 3.5 Sonnet} outperformed \texttt{AgentScope}, suggesting that properly fine-tuned open-source models can enhance the effectiveness of solutions built around closed-source LLMs.

\paragraph{End-to-End Open Solutions}
\label{sec:results:end2endopeness}

Beyond the openness of the LLMs and the public availability of fine-tuned models, some submitters have also made their full solution code publicly available (e.g., agent scaffolds).
The open-source approaches, with at least one submission exclusively using open-source models, are 16, including 
\begin{enumerate*}[label=(\alph*)]
\item  \approach{Alibaba Lingma Agent} \cite{ma2025alibabalingmaagent},
\item \approach{Co-PatcheR} \cite{tang2025copatchercollaborativesoftwarepatching},
\item \approach{DARS Agent} \cite{aggarwal2025darsdynamicactionresampling},
\item \approach{DeepSWE} \cite{deepswe2025},
\item \approach{MCTS-Refine} \cite{wang2025mctsrefinedcothighqualityfinetuning},
\item \approach{OpenHands} \cite{wang2024openhandsopenplatformai},
\item \approach{SWE-Fixer} \cite{xie2025SWEFixerTrainingopensourcellms},
\item \approach{SWE-agent} \cite{yang2024sweagentagent}, 
\item \approach{Skywork-SWE} \cite{zeng2025skyworkswe}, 
\item \approach{mini-SWE-Agent}, and
\item \approach{SWE-Exp} \cite{chen2025sweexpexperiencedrivensoftwareissue}.
\end{enumerate*}
This level of openness means that, given appropriate infrastructure to deploy a LLM, others can fully replicate and build upon these systems, promoting transparency, reproducibility, and further research.

\begin{shaded}
    
\underline{\bf{Answer to RQ 5 (LLMs):}}
Submissions using proprietary models, particularly the Claude family and more recently Claude 4 Sonnet, achieve the highest precision on both leaderboards.
Solutions that combine closed and open-source LLMs—including those that incorporate fine-tuned models—also demonstrate competitive performance.
Approaches based solely on fine-tuned open-source LLMs tend to yield more modest precision but offer benefits in terms of transparency, reproducibility and adaptability. 
\end{shaded}


\section{Discussion and Implications}
\label{sec:discussion}

\subsubsection*{\bf Patch Overfitting on \swebench{}}

An overfitting patch is a fix that passes all available test cases, including the bug-revealing one, yet remains incorrect due to limitations of the validation oracle (the test suite). 
This issue is well known in APR research from the software engineering (SE) community~\cite{smith2015cure,qi2015analysis, martinez2017automatic}.
Recent studies show that overfitting patches on \swebench{} are frequent~\cite{wang2025solvedSWEBench}. 
Academic submissions linked to the SE community (e.g., \cite{xia2024agentlessdemystifyingllmbasedsoftware,ruan2024specrovercodeintentextraction,Zhang2024AutoCoderRover,ma2025alibabalingmaagent}) typically distinguish correct from overfitting patches.  
In contrast, works targeting the AI community (e.g., \cite{yang2024sweagentagent,li2025patchpilot,zainullina2025guidedsearchstrategies}) report only test-passing results. 
This shows that there is a misalignment between the evaluation methodologies from SE and AI communities.

\underline{Implications for Researchers}:
The problem of patch overfitting is still rarely acknowledged outside the SE community, as it is seldom mentioned in papers from the ML/AI community or in industrial blog posts. 
This calls for greater dissemination by the SE community, for instance by presenting the problem in interdisciplinary venues.
Moreover, researchers are encouraged not only to advance methods that mitigate overfitting but also to ensure that their work results in usable, open-source artifacts.

\underline{Implications for Practitioners}:
Practitioners should be cautious when interpreting resolution rates reported on \swebench{}, as they may be inflated by overfitting patches that pass tests but remain incorrect. Relying solely on leaderboard scores can therefore create unrealistic expectations.

\underline{Implications for Leaderboard and Benchmark Builders}:
The SWE-Bench framework and leaderboards provide a valuable foundation for progress tracking in automated issue repair. 
However, they currently lack mechanisms for correctness validation beyond passing test cases, highlighting the need for more rigorous evaluation practices to complement test-based validation.
Future work could enhance the \swebench{} and other similar platforms (e.g. \cite{Silva2025RepairBench}) by incorporating procedures for post-submission correctness assessment and by enabling the integration of patch validation tools (e.g., as plug-ins) into the evaluation pipeline.
Furthermore, mechanisms allowing patches to be uploaded and accessed directly through the leaderboard would enhance transparency and facilitate more in-depth analysis.





\subsubsection*{\bf Metadata Requirements: Balancing Transparency and Participation.}

\swebench{} leaderboards do not require submitters to provide details about their solutions; for instance, 18 entries (13\%) in \shortverif{} do not report the LLMs used. More detailed metadata would improve transparency, especially for academia, but could also create barriers for submitters and discourage industry participation.
One possible interpretation is that lower metadata requirements are associated with reduced entry friction and are consistent with the strong industry participation observed in \shortverif{} (74 of 133 entries).
This reveals a central trade-off in benchmark design between transparency and participation.

\underline{Implications for Benchmark and Leaderboard Builders:} 

They should balance transparency and participation by defining lightweight but informative metadata requirements that help to understand how results are achieved.

\underline{Implications for  Submitters:} 
They are encouraged to share detailed metadata whenever feasible, as this strengthens transparency even beyond the minimum requirements.

\subsubsection*{\bf Data Contamination}
Some \swebench{} issues predate the training cutoffs of several LLMs, raising the risk of data leakage, with high scores potentially inflated by memorization rather than genuine problem-solving ability \cite{liang2025swebenchillusionstateoftheartllms}. 
This  can create false expectations about the actual ability of  approaches to automate issue repair tasks
This issue also affects other benchmarks such as Defects4J \cite{ramos2025LLMsMemorization}.

\underline{Implications for Practitioners}: 
They should treat leaderboards' results with caution and produce further validations (on other benchmarks or in their own context). 

\underline{Implications for Submitters}:
Submitters are encouraged to document mitigation strategies for potential data leakage. 
Researchers, in particular, should explicitly acknowledge these threats to validity in their publications, contribute to efforts aimed at overcoming them (e.g., by reporting cutoff dates), and consider complementary benchmarks that better control for training cutoff overlap.

\underline{Implications for Researchers}: 
They are encouraged to tackle the data contamination problem through different approaches, such as evolving existing infrastructure—as demonstrated by Zhang et al.~\cite{zhang2025swebenchgoeslive} with \texttt{SWE-bench-Live}, an automated pipeline that continuously updates the benchmark with recent data—or by developing new evaluation platforms for issue repair that are explicitly designed to avoid this problem.

\subsubsection*{\bf Limited Real-World Representation}
\swebench{} covers only a few Python projects. As a result, performance may not generalize to diverse, noisy, real-world scenarios and could be overestimated, as suggested by Zeng et al.~\cite{zeng2025skyworkswe}, who reported much lower resolve rates on their benchmark of 10,169 Python tasks.

\underline{Implications for Submitters}: They should avoid overstating generalization when reporting results on \swebench{}, as strong performance may not transfer to broader, real-world scenarios. This is important for industry submitters, where leaderboard results are sometimes highlighted in publicity or product communications.

\underline{Implications for Researchers and Practitioners}: Evaluate on other benchmarks to validate whether reported performance generalizes beyond \swebench{}.

\subsubsection*{\bf Emerging Variants of \swebench{}}






Sub-benchmarks from \swebench{} have emerged.
For instance, Xia et al.~\cite{xia2024agentlessdemystifyingllmbasedsoftware} constructed \texttt{SWE-bench Lite-S} by filtering out problematic issues.
Ma et al. proposed \texttt{SWE-bench-Lite-FIX}~\cite{OLDma2024understandRepoUnderstander}, an evolution of  \shortlite{} where 45 issues were improved by augmenting the issue description.
Creating a subset of a benchmark could be benefit for showing the real performance of one approach, but, at the same time, may complicate further comparison with other approaches evaluated on the original benchmark.
Without controlled versioning and official support -such as recognition or integration by the leaderboard maintainers, as happened with \shortverif{}- these new subsets or evolutions of existing ones may remain underutilized and have limited influence within the research community. Currently, \swebench{} does not  track submissions on emerging forks and derivatives as the mentioned ones. 

\underline{Implications for Leaderboards and Benchmarks Builders}:
Encouraged to provide mechanisms for the systematic registration and tracking of benchmark subsets.

\underline{Implications for Researchers}:  Encouraged to establish dialogue with leaderboard maintainers to ensure visibility and comparability of their new subsets, and to foster the development of mechanisms for their systematic registration and tracking.

\subsubsection*{\bf Benchmark versioning}

The evolution of benchmarks often aims to improve data quality or address inconsistencies (such as those pointed by~\cite{Aleithan2025Revisingswebench}). This trend is not unique to \swebench{}; for example, other bug benchmarks such as Defects4J \cite{Just2014Defects4J} have undergone significant revisions that include issues deprecation.
To the best of our knowledge, \swebench{} does not yet provide explicit versioning.

\underline{Implications for Leaderboard and Benchmark Builders}:
Encouraged to provide mechanisms for explicit versioning and to relate results across versions, along with clear documentation of the rationale behind revisions, in order to preserve traceability and comparability.

\subsubsection*{\bf Saturation and the Cherry-Picking of Submissions}

In September 2025, submissions to \swebench{} achieved up to 76.8\% precision—an increase of more than 20 percentage points compared to the previous year.
This progress coincides with the release of increasingly capable LLMs. 
As shown in Table~\ref{tab:results:llmCombinationOneColumn}, all systems exceeding 70\% precision relied on state-of-the-art Claude 4 models, either alone or in combination with others.
If such trends persist,  upcoming LLM releases could surpass 80\% or even  90\%, suggesting that \swebench{} may be approaching a saturation point.
This perception of saturation may also be reinforced by selective reporting when using different LLMs: submissions may test several LLMs but only the best outcome is submitted, hiding result variability. 

\underline{Implications for Submitters (Researchers, Industrials)}:
Encouraged to disclose not only peak results but also the range of performance observed across different LLMs. An illustrative example is the submission from Princeton University, which reported 21 entries for \approach{mini-SWE-Agent}, each using a different LLM. Such transparency helps clarify whether apparent saturation reflects consistent LLM ability or selective reporting. 
Moreover, as frontier LLMs converge toward high precision, researchers should explore extended tasks or new benchmarks.

\underline{Implications for Leaderboard Builders}:
Encouraged to provide mechanisms to register the variability of approaches across the LLMs used in each entry and to display aggregate statistics.

\subsubsection*{\bf Cost Barriers: Cloud-Based and Fine-Tuned LLMs}
As shown in Table~\ref{tab:results:llmCombinationOneColumn}, most of the LLMs used—and nearly all top-performing ones—are cloud-based LLMs accessed via an API. Access to these LLMs incurs monetary costs, which creates a barrier to experimentation, particularly in academic settings where budgets are more constrained. Although some solutions leverage open-source LLMs, many applied fine-tuning, which is itself computationally expensive and therefore also subject to cost limitations.

\underline{Implications for Submitters}:
Encouraged to also report results obtained with open-source LLMs, even if these do not achieve top precision. Such submissions increase diversity on the leaderboard, promote transparency, and provide baselines for the community.

\underline{Implications for Practitioners}:
They should be aware that leaderboard performance often comes at a monetary cost, not always explicit.
This highlights the need to carefully evaluate cost-performance trade-offs before adopting a solution in practice.

\underline{Implications for Benchmark Builders}:
Encouraged to consider costs as relevant dimensions of evaluation, which may include financial, computational, and energy costs. 

\section{Threats to Validity}
\label{sec:ttv}

\paragraph{\textbf{External Validity.}}
In this study, we focused on two \swebench{} leaderboards, which we selected due to the SWE-bench's substantial impact on both academia and industry, as demonstrated both by the companies involved (most industry big players) and the scientific impact, e.g. with  the seminal paper \cite{jimenez2024SWEBenchLLMs} reaching over 950 citations on Google Scholar as of September 2025. Other benchmarks may be equally representative of the broader issue-fixing landscape, but we do not claim that our findings can be applied to them. 

\paragraph{\textbf{Internal Validity.}} We followed a systematic procedure to collect information from the leaderboard submissions. However, there is a risk that we may have missed artifacts describing some entries, resulting in certain submissions being left uncharacterized.
To mitigate this, we extended our search beyond the official leaderboard metadata, consulting multiple sources including Google searches, LinkedIn profiles, and scientific publications indexed on Google Scholar and arXiv.
Nevertheless, there remains a risk that the reported information does not accurately reflect the actual systems used.


\paragraph{\textbf{Construct Validity.}}

Some potentially important dimensions and variables were not analyzed in this study.
First, we did not account for the number of parameters or the specific minor version of each LLM. 
This decision was made to simplify the analysis and focus on identifying broader trends across model families and major releases, rather than fine-grained differences between specific versions.
Second, we did not evaluate the monetary cost associated with the different submissions, despite some submitters reporting this information in the metadata.
While cost is a relevant factor for assessing the usability and practicality of an approach, it was excluded from our analysis for practical reasons.
Token prices have generally decreased over time, making cost comparisons across submissions potentially misleading. 

\section{Related work}
\label{sec:relatedwork}

Previous work have studied in details the effectiveness of approaches evaluated on \swebench{} and published on the leaderboards.
For example, 
Meng et al. \cite{meng2024empiricalstudyllmbasedagents} conducted an empirical study of the efficiency of 7 repair systems (including \cite{liu2024marscode,Zhang2024AutoCoderRover,ouyang2024repographenhancingaisoftware,xia2024agentlessdemystifyingllmbasedsoftware}) on \shortlite{}.
They found that issue quality significantly influences the effectiveness of resolution methods.
Aleithan et al.~\cite{Aleithan2025Revisingswebench} presents a manual analysis of 251 patches generated by the \approach{SWE-Agent + GPT-4} submission. They  found that in 32.67\% of the cases, the issue description included the complete solution, effectively leaking the correct patch. Additionally, 12.75\% of the patches were incorrect yet still passed the test suite.
Wang et al. \cite{wang2025solvedSWEBench} presents an empirical study on the correctness of plausible patches generated by three systems evaluated on \sweverif{}. The study reveals a flaw in the patch validation process of \swebench{} causing that, on average, 7.8\% of plausible patches were actually incorrect.
Bouzenia et Pradel \cite{bouzenia2025understanding} studied the trajectories of three agents, two  with results submitted to \swebench{}.
They found for instance,  that successful trajectories (that conduct to correct patches) balance exploration, explanation, and validation steps, while unsuccessful ones exhibit repetitive, non-adaptive action cycles.
Similarly, Ceka et al.\cite{ceka2025understandingsoftwareengineeringagents} conducted an empirical study of  trajectories submitted to \sweverif{}, while Chen et al.\cite{chen2025unveilingpitfallsunderstandingaidriven} analyzed complete trajectories and testing logs from eight agents on the same benchmark 
Liang et al.~\cite{liang2025swebenchillusionstateoftheartllms} investigate whether the performance of LLM-based approaches on \swebench{} is driven by genuine coding capabilities or by memorization.  
Their findings show that models such as {OpenAI o3-mini}  are still able to repair the majority of issues but performance drops when evaluating issues outside of that benchmark. 


\section{Conclusions}
\label{sec:conclusion}

In this paper, we presented the first comprehensive study of the submissions to the SWE-Bench Lite and SWE-Bench Verified leaderboards. 
We analyzed \totalentries{} entries from \totalsubmitters{} submitters, examining their origins, product characteristics, and LLM usage.
Our results showed that the majority of high-performing submissions relied on proprietary LLMs, particularly Claude 4.
We also found that most submissions came from industry, including both large tech companies and small startups, as well as independent developers. 
By systematically analyzing the submissions to the \swebench{} leaderboards, we uncovered the range of profiles behind them, contributing to a deeper understanding of the current state of the field.

\begin{acks}
This paper has been funded by the “Ramon y Cajal” Fellowship (RYC2021-031523-I), by grant PID2024-156019OB-I00 funded by MICIU/AEI/10.13039/501100011033 and by ERDF, EU. 
\end{acks}

\bibliographystyle{plain}
\bibliography{references}

\end{document}